\documentclass[12pt,preprint]{aastex}

\newcommand{\chandra}{{\sl Chandra\/}}
\newcommand{\asca}{{\sl ASCA\/}}

\shorttitle{Luminous X-ray Binaries in M101}
\shortauthors{Mukai et al.}

\begin{document}


\title{Chandra Observation of Luminous and Ultraluminous X-ray Binaries in M101}


\author{K. Mukai\altaffilmark{1}, W. D. Pence, and S. L. Snowden\altaffilmark{1}}
\affil{Code 662, NASA/Goddard Space Flight Center, Greenbelt, MD 20771}

\and

\author{K. D. Kuntz}
\affil{Joint Center for Astrophysics, Physics Department,
	 University of Maryland Baltimore County, 1000 Hilltop Circle,
	 Baltimore, MD 21250, USA.}


\altaffiltext{1}{Also Universities Space Research Association}


\begin{abstract}

X-ray binaries in the Milky Way are among the brightest objects
on the X-ray sky.  With the increasing sensitivity of recent missions,
it is now possible to study X-ray binaries in nearby galaxies.  We present
data on six luminous sources in the nearby spiral galaxy,
M101, obtained with the \chandra\ ACIS-S.  Of these, five appear to
be similar to ultraluminous sources in other galaxies, while the
brightest source, P098, shows some unique characteristics.
We present our interpretation of the data in terms of an optically
thick outflow, and discuss implications.

\end{abstract}


\keywords{galaxies: X-rays---galaxies:individual (M101)---galaxies: spiral}


\section{Introduction}

X-ray binaries in the Milky Way, with typical intrinsic luminosities
in the range 10$^{34}$--10$^{38}$ ergs\,s$^{-1}$ at a typical distance
of 8 kpc, dominate our 2--10 keV sky.  We therefore know a great deal
about these Galactic X-ray binaries through many studies over the
last several decades (see \citealt{WNP95} for a review).  They are close
binaries in which a neutron star or a black hole is accreting from a
non-degenerate companion.  They can be divided into low-mass X-ray
binaries (LMXBs) and high-mass X-ray binaries (HMXBs) depending on
the spectral type of the mass donor.  The HMXBs are young objects
that are concentrated in the Galactic plane, preferentially in spiral arms. 
The LMXBs, on the other hand, appear to belong variously to the old disk,
bulge, and globular clusters.  Many neutron star LMXBs show
thermonuclear flashes (i.e., type I bursts) suggesting a relatively
low-magnetic field; many HMXBs are coherent X-ray pulsars, containing
highly magnetized neutron stars.

The Eddington limit for a 1.4 M$_\odot$ object is $\sim 2 \times 10^{38}$
ergs\,s$^{-1}$.  Significantly more luminous X-ray binaries can be considered
black hole candidates on this argument alone, although HMXB pulsars SMC X-1
and LMC X-4 appear to exceed this limit at times (however, we do not know
if they actually violate the Eddington limit, since the emission geometry
of X-ray pulsars is not spherically symmetric).  The definitive evidence
for a black hole in X-ray binaries comes from radial velocity studies of
the mass-donor in the optical.  In particular, over a dozen soft X-ray
transients (SXTs; a subtype of LMXBs) have a measured mass function which
exceeds 3 M$_\odot$, with inferred
compact object masses typically in the 5--15 M$_\odot$ range \citep{BO98}.
In the X-ray regime, black hole binaries have characteristic spectral shapes,
including a low/hard state dominated by a power law spectrum, and
a high/soft state dominated by a $\sim$1 keV thermal
component, usually interpreted as arising from the inner disk \citep{T95}.
Although detailed studies have led to suggestions of additional spectral
states \citep[see, for example,][]{Z01}, these spectral states are different
from those of typical neutron star systems, either a 5--10 keV
bremsstrahlung-like spectrum (LMXBs) or a power-law with an exponential
cut-off (HMXBs).

Although a great deal is known about Galactic X-ray binaries,
studies of extragalactic X-ray binaries offer complementary
insights.  In particular, a complete census of Galactic systems
is difficult due to the extinction in the Galactic plane, and
the luminosity estimates of Galactic systems are generally subject
to large uncertainties.  Study of a face-on galaxy such as M101
allows a study of a luminosity-limited sample with, on average,
a low absorbing column.

In recent years, we have gained an additional motivation to
study extragalactic X-ray source populations, in the form of off-nuclear
point sources with luminosities greater than
10$^{39}$ ergs\,s$^{-1}$ (hereafter Ultraluminous X-ray sources,
or ULXs\footnote{Sources with luminosities in the 10$^{38}$ -- 10$^{39}$
ergs\,s$^{-1}$ range have also been considered ULXs by some authors.
They are likely to be related to ULXs, as well as to Galactic
BHCs.}) that have been found in many nearby galaxies \citep{C99}.
One possible interpretation is that they are accreting intermediate
mass (10$^2$--10$^4$ M$_\odot$) black holes.  This would be exciting
if confirmed, because previously known black holes could be categorized
into stellar ($\leq$10 M$_\odot$) or supermassive ($> 10^6$ M$_\odot$)
subclasses. However, disk blackbody models of the \asca\ spectra of ULXs
\citep{M00} suggest they may have accretion disks as hot as several
keV at their inner edges.   In the standard model, it is difficult
for a disk around an intermediate mass black hole to achieve such
high temperatures.

The superb angular resolution of \chandra\ allows the detection
of point sources well below 10$^{38}$ ergs\,s$^{-1}$ in nearby
(say, closer than 10 Mpc) galaxies, thus sampling LMXBs,
HMXBs, and both black hole and neutron star systems.  Consequently,
many groups are studying X-ray source populations in a considerable
number of nearby galaxies as summarized, for example in \citet{Pr01}.
Here we present preliminary results of our observation of M101.

\section{Observation and source detection}

M101 is a nearby face-on spiral galaxy at an estimated distance
of $\sim$7.2 Mpc.  It is an ideal galaxy for the observation
of X-ray binaries, supernova/hypernova remnants \citep{S01},
and diffuse emission \citep{K02}.
We have therefore observed M101 with \chandra\ ACIS for 98.2 ksec
during 2000 March 26--27.  The details of the observation, data reduction,
source search, the catalog of 110 sources detected on the S3 chip
and their collective properties are described in \cite{P01}.
Here we concentrate on the 6 brightest sources, listed in Table~1.
We adopt the source number in \citet{P01} as their names throughout
this paper.  \cite{P01} argue on statistical grounds that about 75\%
of the 110 sources detected on the S3 chip are intrinsic to M101.
Therefore, we will assume that the majority of the six sources in
question, if not all, are located in M101.

One of the sources, P098, is an order of magnitude brighter (as seen
with ACIS-S) than the others and has other peculiar properties;
this object will be the focus of this paper.  However, we will first
discuss the other 5 systems listed in Table~1 with observed
luminosities in excess of 10$^{38}$ ergs\,s$^{-1}$.  The inferred
bolometric luminosities are higher, almost certainly exceeding the
Eddington limit for a 1.4 M$_\odot$ object.  These 5 systems
(three of which have been discussed in \citealt{S01} as likely
binaries, rather than hypernova remnants)
are therefore black-hole candidates (BHCs), even though they
do not qualify as ULXs using the threshold of 10$^{39}$ ergs\,s$^{-1}$.

\section{The Five Bright Black-Hole Candidates}

We have attempted a simple continuum spectral fit to the five BHCs,
using power-law, bremsstrahlung, blackbody and disk blackbody (diskBB,
as implemented in XSPEC)
models.  The spectrum of P104 is best fit with a power-law model, while
the diskBB model works best for the others.  The best-fit models do not
resemble typical spectra of Galactic X-ray pulsars (flat power-law) or
of bright neutron star LMXBs (Comptonized spectra which can be approximated
as a 5 keV bremsstrahlung; \citealt{WNP95}).  We have also examined the
timing properties of these sources: their light curves in 5000 s bins
are shown in Fig.~1.  Of the five, P104, the power-law source, is highly
variable on a relatively short timescale (e.g., note the factor of
$\sim$2 drop in count rate in one 5000 s bin near the end
of the observation).  This is also one of the hypernova
candidates of \citet{W99a}, since it is coincident with a optically detected
supernova remnant MF83 \citep{M97}, but the observed
variability excludes the hypernova interpretation \citep{S01}.
The power-law index of P104, $\sim$2.6, is unusually steep for a black
hole binary in a low/hard state.   Although such an index is often seen
in the power-law component of high/soft state
\citep[e.g., 2.2--2.7 in GS~1124$-$68:][]{E94},
a soft component is not obvious in the \chandra\ spectrum
of P104: A single power-law model fit ($\chi^2_\nu$=1.1) is clearly
favored over a diskBB model fit ($\chi^2_\nu$=2.2) and
also over a power-law plus disk blackbody model fit ($\chi^2_\nu$=1.2), though
the last is extremely sensitive to initial guesses and thus not reliable.
Nevertheless, the rapid variability clearly establishes P104
as an X-ray binary in M101.

According to a $\chi^2$ test for constancy of the light curves of
these 5 sources, only P104 is found to be significantly variable,
in the total \chandra\ band.  Of the remaining 4 sources, P76 is
a possible exception: when its light curve in the 2--8 keV band
is tested for constancy, we obtain $\chi^2$=24.9 for 19 degrees
of freedom.  The probability of a constant source displaying this
level of apparent variability is $\sim$16\% (in the total \chandra\ band,
we obtain $\chi^2$=16.1, and a chance probability of 65\% for P076).

We also detect an apparent emission line in the spectrum of P104
at 1.02 keV, probably the Ne X Ly$\alpha$ line (Fig.~2).  This line
may be intrinsic to the binary, since it is present in one Galactic
X-ray binary, 4U~1626$-$67 \citep{A95}.  However, another likely origin
of the line is the hot plasma in MF83.  The total luminosity in this line is
of order 10$^{37}$ ergs\,s$^{-1}$ at the distance of 7.2 Mpc.  Other lines
are not required or excluded by the data.  Similarly, in the spectrum of
P005 there is an apparent line at 1.34 keV with an inferred line luminosity
of $\sim 4 \times 10^{37}$ ergs\,s$^{-1}$, and another possible line
at 1.85 keV.  It is interesting to note that P005 is one of the eight
``interarm'' sources detected in the S3 chip \cite{P01}, which coincides
with the ROSAT HRI source H18 of \cite{W99b}.
The latter authors suggested a blue optical counterpart, interpreted as
AGN; however, the \chandra\ and optical positions are about 5 arcsec apart
\citep[][and Wang, private communication]{W99b},
and therefore this source may turn out to be in M101 after all.
The analogy with P104 suggests a combination of a luminous X-ray binary,
responsible for the optically thick continuum, and a supernova remnant,
responsible for the line emission.

The spectra of the three remaining BHCs (P076, P070, and P110) can
be characterized as a disk blackbody with inferred temperatures at
the inner edge of the disk in the 0.6--1.6 keV range, with no
obvious emission lines.  Their luminosities in the \chandra\ band
are inferred to be 1.7--4.0 $\times 10^{38}$ ergs\,s$^{-1}$.
Together with P104, these 4 sources appear to be accreting black holes
in high/soft states, similar to the ULXs and other bright BHCs observed
with \asca\ \citep{M00}.  For the interpretation of the diskBB model
parameters as disk inner radius and temperature to be viable,
P076 and P005 need to be relatively low-mass ($\leq$3 M$_\odot$)
black holes accreting at the Eddington limit.

\section{The Peculiar ULX P098}

The brightest source we have discovered in M101, P098, stands out from
others in luminosity, variability, and spectral shape.
We have already noted the clear variability in P098 \citep{P01}.
Furthermore, the variability is far more pronounced in the
0.8--2.0 keV range than in the lower energy ranges (Fig.~3).
We have extracted spectra of P098 at three time intervals
(indicated in Fig.~3) and fitted them with simple models (Fig.~4).
Even during interval (a), when P098 had the highest count rate and
the hardest spectrum, most of the flux is in the soft band (E$<$1 keV).
It is clear that LMXB-like ($\sim$5 keV bremsstrahlung) or HMXB pulsar-like
(flat power law with photon index $\sim$1.5) models are inappropriate.
In fact, we find that no single-component model fits the interval (a)
spectrum satisfactorily over the entire ACIS-S band.  The diskBB model
did best with $\chi^2_\nu \sim 2.0$, which fits the data below $\sim$1.5 keV
adequately but leaves residuals above 2 keV.

When we restrict ourselves to fitting data below 1.5 keV only,
either blackbody or diskBB models work well for the bright state,
giving a reduced $\chi^2$ of 1.1 (for blackbody)
and 1.3 (for diskBB) in this range.  The high energy excess can then be
fit using an additional power-law component.   In a similar attempt at
fitting the spectra for intervals (b) and (c), we find no formally
acceptable models; however, again, the blackbody and diskBB models
provide the ``best'' fits, with $\chi^2_\nu \sim 2$.  The residual
patterns are complex and are not dominated by obvious features,
although a high energy excess component can reduce $\chi^2_\nu$.
We therefore use the blackbody and diskBB models to characterize
the spectral changes from low count rate (b) to intermediate (c)
to high (a) count rate states of P098.

Using the blackbody model, the inferred temperature ranged from
0.09--0.17 keV, the radius changed from $\sim$20,000 km to $\sim$5,000 km,
while the bolometric luminosity (calculated analytically from the model
normalization and the distance to M101 of 7.2M pc, assuming all absorptions
are external to the X-ray emission region) stayed near
$\sim 3 \times 10^{39}$ ergs\,s$^{-1}$ (see Table~2).
This is in contrast to the behavior of the luminosity in the 0.2--1.5 keV
band, which changed by nearly a factor of 3.
With the diskBB model, the temperature range was 0.1--0.2 keV,
the inner radius of the disk changed from $\sim$20,000 km to $\sim$4,000 km,
while the bolometric luminosity (from model parameters) stayed at
$\sim 5 \times 10^{39}$ ergs\,s$^{-1}$.
The differences between parameters derived using these two different models
give some indications of true uncertainties in derived parameters.
Using either model, the inferred radii are anti-correlated with the
temperatures, and are far larger than the typical inner disk radius (Table~1).

The apparent variability in the \chandra\ band can be described as changes
in high-energy cut-off of the spectrum, while the count rates in the 0.2--0.5
keV band changed little.  This cannot be explained by a cold absorber;
we also consider a highly ionized absorber to be unlikely, as we do not
see warm absorber edges such as due to O VII and OVIII.  The variability
above 0.8 keV may be due to appearance and disappearance of a hot continuum
source.  Since such a continuum source also contributes lower energy
(0.2--0.5 keV) flux as well, there must be compensating change in the
lower temperature continuum source to keep the low energy counts unchanged.
This is exactly what the diskBB fits imply.  An alternative interpretation
is a simple change of temperature of the emission region, as parameterized
by the blackbody fits.  Therefore, even though neither model provides
a statistically satisfactory fits for intervals (b) and (c), they are
likely to reflect the kind of changes taking place in P098.  We therefore
consider the implications of the blackbody and diskBB model fits.

In many other ULXs, we have a puzzle in that the black hole mass
inferred from the Eddington limit argument is high, while that inferred
from using the disk blackbody model is low.  This is particularly severe
if Schwarzschild black hole is assumed; even the assumption of Kerr black
holes may not completely solve this puzzle \citep{E01}.  In P098, however,
the situation is very different.  The low temperature and the large
luminosity can both be accommodated in the framework of an intermediate
mass black hole accreting at much less than the Eddington rate.
The problem with this picture is the large disk radius changes inferred
by the fit (roughly by a factor of 4) while the inferred bolometric
luminosity changes little.  It is difficult to understand how a slight
change in the mass accretion rate (as suggested by the near-constant
luminosity) can trigger such a drastic change in inner radius of
the accretion disk in such a short timescale.  Looking at this from
another angle: since a standard accretion disk has an $T \propto R^{-3/4}$
profile, the temperature at $R=20000$\,km should be 0.3 times that at
$R=4000$\,km at any given moment, while the ratio of inferred temperatures
at the (moving) inner edge of the disk is 0.5.  Thus, if take the diskBB
model fits at face value, P098 must have lost the inner part of the disk,
while the temperature at $R=20000$\,km increased by 0.5/0.3$\sim$1.67, or
the luminosity of the remaining part of the disk by a factor of $\sim$8.
This seems rather contrived for any potential models in which the
blackbody-like component originates in the accretion disk; we therefore
prefer an alternative interpretation.

The behavior of P098 is reminiscent of the slow evolution of
classical novae in the constant bolometric luminosity phase \citep{B98}
and of the X-ray variability of super-soft sources \citep{S96}.
In these systems, nuclear burning on the surface of an accreting
white dwarf keeps the radiative luminosity at or near the Eddington limit.
The nuclear energy also drives a strong outflow; this wind is optically thick,
hence the observed spectrum is determined by the radius of last
scattering.  When the outflow rate is higher, the effective
photospheric radius is large, hence the observed temperature is
low.  When the outflow rate is lower, the photosphere is smaller,
hence a higher temperature is seen.  The situation in P098 may be analogous.

The relative lack of variability in bolometric luminosity of P098
suggests the existence of a limiting mechanism: we postulate that
this is the Eddington limit, therefore implying a black hole mass of
$\sim$15--25 M$_\odot$.  The Eddington limit corresponds to an accretion
rate of a few times 10$^{19}$ g\,s$^{-1}$; if the accretion disk receives
material from the mass donor at a higher rate, the excess will likely be
ejected from the system, keeping the radiative luminosity near the
Eddington limit.  Since the study of the resulting outflow (jets and/or
winds) is a vast topic with no clear-cut answer as of yet, and because
there is likely to be strong viewing-angle dependence, we can only
make simple order-of-magnitude estimates.  We start with the mass
continuity equation, $4\pi r^2 V(r) \rho(r) = \dot M$, and integrate
the density $\rho(r)$ from inner boundary $R$ to infinity to estimate
the column density down to $R$, $N_H(R)$.  Since the velocity law $V(r)$
is unknown, we adopt the simplest assumption of a constant velocity,
which results in $N_H(R)=\dot M/4\pi VR$.  For fiducial values of
$R=10000$ km (within the range of size inferred by fitting blackbody
and diskBB models to the spectra of P098), $\dot M = 1.0 \times 10^{19}$
g\,s$^{-1}$ (comparable to the Eddington accretion rate), and
$V=10000$ km\,s$^{-1}$, we obtain $N_H = 0.77$ g\,cm$^{-2}$.  This implies
that an electron scattering opacity of order unity is possible.  The fiducial
velocity we have adopted is in between that observed in the X-ray P Cygni
profile of the neutron star system, Cir X-1 \citep{B00} of 2000 km\,s$^{-1}$
and the escape velocity at 100 Schwarzschild radii ($R \sim 6000$ km).
Thus, even though the assumption of a constant velocity outflow is suspect,
we believe our simple calculations are sufficient to demonstrate an
order-of-magnitude feasibility of our interpretation.

Within this framework, we observe X-ray photons scattered in the mass
outflow from P098, at a typical radius of 10000 km.  Fluctuations in
the mass loss rate and/or the outflow velocity result in varying radius
of this effective photosphere, while the total radiative luminosity
remains constant at about the Eddington limit.  The changing radius
of the effective photosphere drives the correlated change in the
temperature and model normalization.
It is possible that such a mechanism is responsible for
most of the variability of P098 down to the fastest timescale
detected (Fig.~3), not just the spectral changes between intervals
{\sl a\/}, {\sl b\/}, and {\sl c\/}, though we cannot prove
this with the available data.  If the dominant source of the variability
is mass loss rate rather than the velocity, it is likely that higher
accretion rate leads to higher mass loss rate, a larger photosphere,
a lower X-ray temperature, and hence a lower count rate with \chandra\ ACIS-S.

A detection of P Cygni profile, particularly in the X-rays
where the mass donor does not contribute a significant flux,
would be a strong evidence for the mass outflow interpretation,
similar to what has been achieved for Cir X-1 \citep{B00};
however, this probably requires a future generation of X-ray
observatory such as Constellation-X.  If the optical counterpart
can be identified and is not dominated by the mass donor, X-ray and
optical brightness of P098 may be anti-correlated.  On the other hand,
if future X-ray observations reveal that the bolometric luminosity is
far from constant, we will be forced to rethink our interpretation.

\section{Implications for other ULXs}

We have studied the six brightest sources detected in our
\chandra\ ACIS-S observation of M101.  We briefly consider
the possible implications of our findings on the nature of
ULXs in general.

Of the 5 non-ULX BHCs that we have studied, one (P104) is
spatially coincident with a supernova remnant (MF83).  Another
source, P005, although located in the interarm region and
originally suspected of being a background AGN, may also be
a combination X-ray binary/supernova remnant in M101.  In our own Galaxy,
the jet source SS~443 is in the supernova remnant W~50 \citep{S76},
while the super-Eddington neutron-star binary Cir X-1 has tentatively
been linked to the nearby supernova remnant G~321.9$-$0.3 \citep{S93}.
The case of P104 may point towards an interesting link between
bright extragalactic BHCs and some of the unique and extreme X-ray
binaries in our Galaxy.

The ULX P098 is highly variable, particularly above 0.8 keV,
and has a soft blackbody-like spectrum.  Fitting of spectra from
P098 extracted from three time intervals shows that the bolometric
luminosity may have been relatively constant.  The apparent
violent variability appears to reflect anti-correlated changes
in the source temperature and size.  We have therefore presented
an interpretation based on optically thick outflow, because we find it
unlikely that the inner radius of the accretion disk can change by
a factor of 4 on such a short timescale, without a commensurate
change in bolometric luminosity.

Yet the spectrum of P098 can be fit with a disk blackbody model and
a power law excess, a traditional model for black hole candidates
which is also applied to ULXs \citep[e.g.,][]{E91, MFR00, E01}.
While our \chandra\ data of P098 can be fit with the diskBB model,
we have questioned the standard interpretation.  It is important
to note that a variety of physical pictures can reproduce a diskBB-like
X-ray spectral shape, provided that the emission region is
optically thick, has a size similar to the inner disk, and has
a slight temperature gradient.  That is, the optically thick disk
interpretation of ULX spectra \cite{M00} is almost certainly not
unique. An additional argument for caution is provided by \citet{Z01},
who show that the soft component in Galactic BHCs are often too broad
to be fit by models of an optically thick disk.
They argue that an intermediate temperature material is likely
present, providing either additional blackbody contributions or
additional Comptonization.  Unfortunately, existing X-ray spectra
of extragalactic ULXs are not sufficiently constraining to allow
discrimination between pure disk models, Comptonized disk models,
and models of X-rays scattered in a strong outflow.

We may be able to sidestep the problem of disks that appear to be too
small and too hot for the black hole mass by abandoning the ``pure disk''
assumption for the ULX spectra.  However, it leaves the question of the black
hole mass unresolved: ULXs are either unbeamed objects containing
intermediate mass black holes, or they are beamed objects containing
stellar-mass black holes.  Our contributions to this debate are twofold.
First is that no object we have detected in M101 requires an intermediate
mass black hole.  Even the most luminous object, P098, can be an
Eddington limited, unbeamed X-ray source with a $\sim$20 M$_\odot$
black hole.  This is somewhat larger than the typical mass of stellar
black holes, but not so large as to require a new class of objects,
at least in M101.

Our second contribution to this debate comes from the strong variability
seen in P098 and P104.  The very fact that variability can be detected
with less than a thousand photons suggests an impressive degree of
variability in P104, perhaps favoring a beamed model for ULXs such as
suggested by \cite{K01}.  In addition, the energy dependent
light curves of P098 suggest the possibility that the true variability
of other ULXs may have been underestimated; similar analysis of
light curves of P076 proved suggestive, but not conclusive.
It would be very important to search for similar energy-dependent
variability characteristics in other ULXs, whenever counting statistics
permit: existing studies may have severely underestimated the true
variability of ULXs if they often are variable predominantly at higher
energies, because observed counts are generally weighted heavily towards
lower energies.



\clearpage


\begin{figure}
\plotone{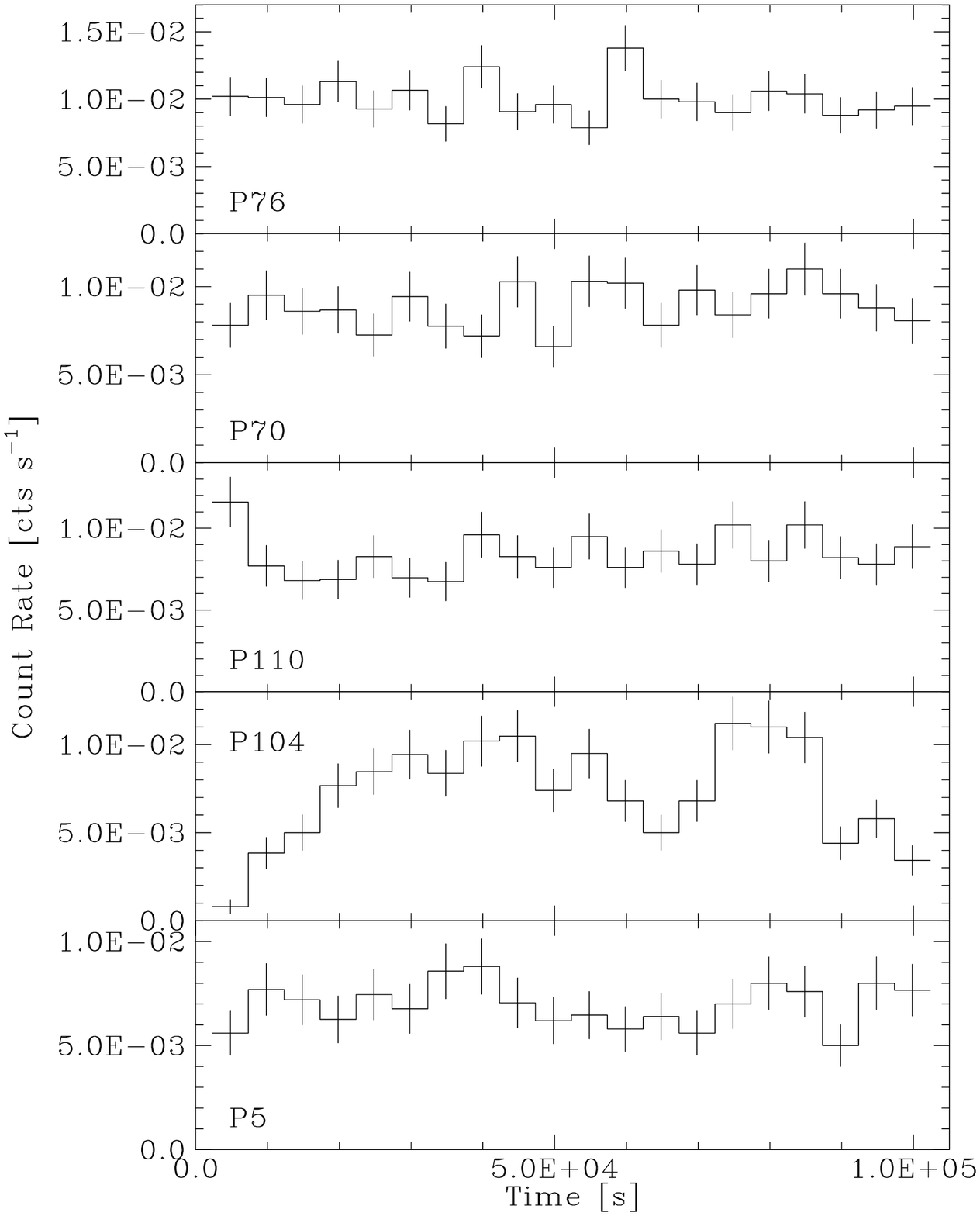}
\caption{\chandra\ ACIS-S (0.125--8.0 keV)
light curves of the 5 bright normal ULXs.}
\end{figure}

\begin{figure}
\plotone{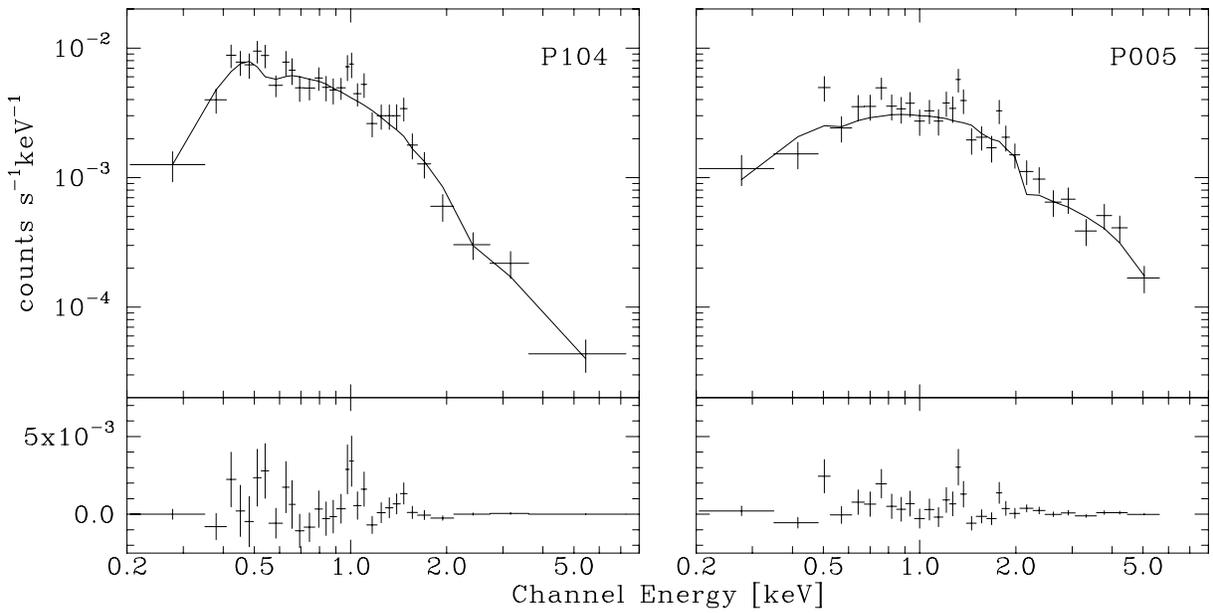}
\caption{\chandra\ ACIS-S spectra of P104 and P005.
Observed spectra with the best-fit continuum (power-law for P104
and disk blackbody for P005) are plotted in the upper panels, while
the residuals are shown in the lower panels.  There are apparent
emission line-like features at 1.02 keV (P104) and 1.34 keV (P005).}
\end{figure}

\begin{figure}
\plotone{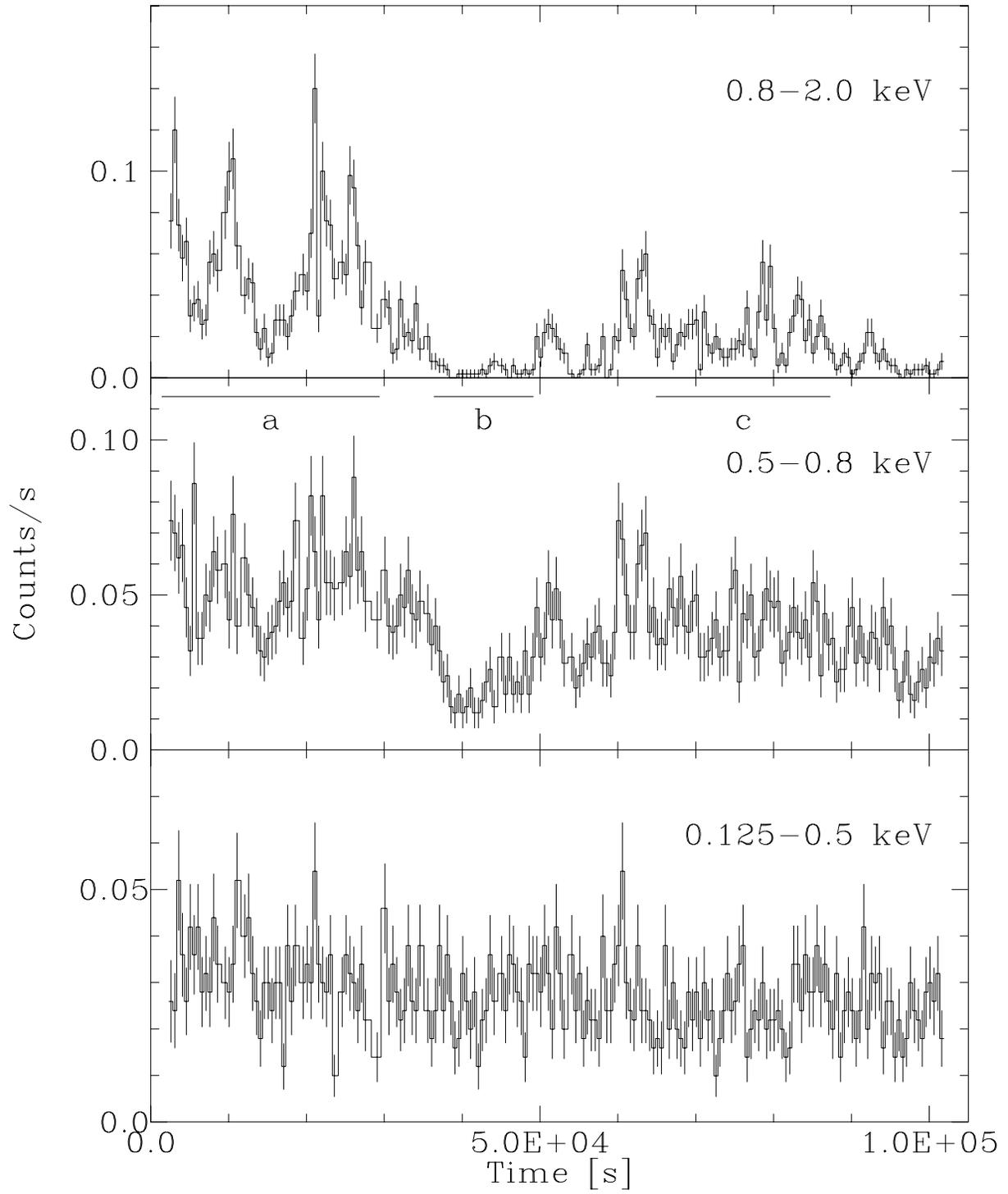}
\caption{\chandra\ ACIS-S light curves of P098 in 3 energy bands.
Horizontal bars in the second panel indicate the time intervals
selected for spectral analysis.}
\end{figure}

\begin{figure}
\plotone{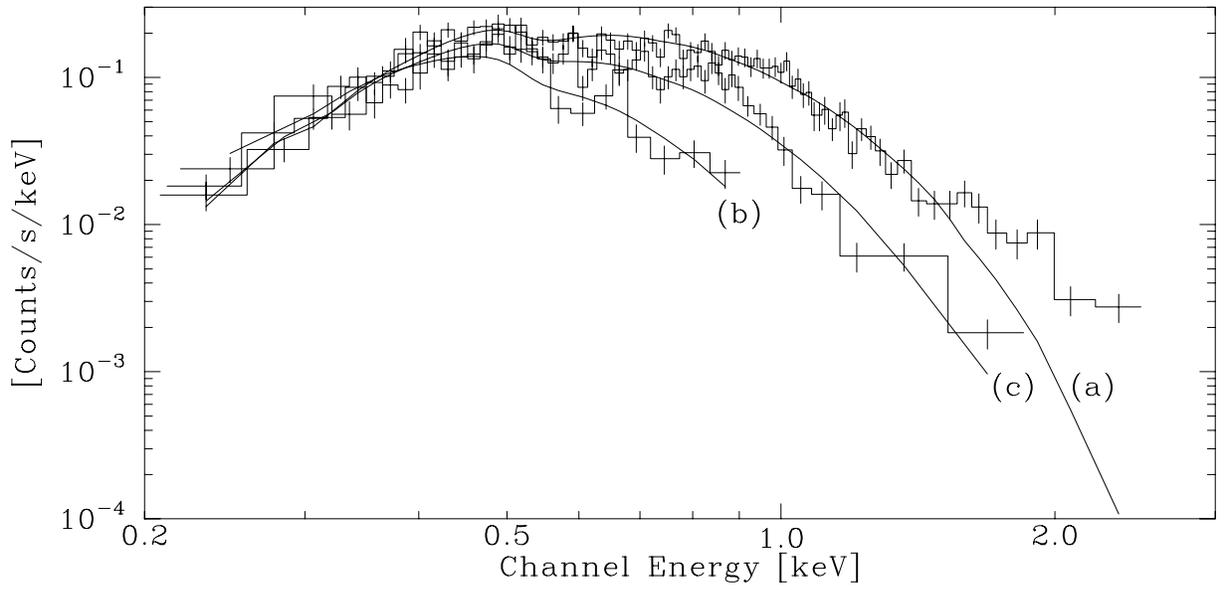}
\caption{\chandra\ ACIS-S spectra of P098 during 3 time intervals
indicated in Fig.~3.  The spectral fits were performed in the restricted
energy range below 1.5 keV, even for interval (a) during which source is
strongly detected above this energy.}
\end{figure}

\clearpage

\begin{deluxetable}{lrrrrr}
\tablecaption{ULX and Black Hole Candidates in M101}
\tablehead{ \colhead{Name} & \colhead{Net counts}
	& \colhead{Spectral Model} & \colhead{T$_{in}$ (keV)/$\alpha$}
	& \colhead{R$_{in}$ (km)} & \colhead{L (ergs\,s$^{-1}$)} }
\startdata
P098 & 9308 & diskBB & 0.18 & 4400.0 & 1.4$\times 10^{39}$ \\
     &      &        &      &       & \\
P076 &  942 & diskBB & 1.61 & 20.2   & 4.0$\times 10^{38}$ \\
P070 &  872 & diskBB & 1.07 & 32.0   & 2.5$\times 10^{38}$ \\
P110 &  777 & diskBB & 0.58 & 89.6   & 1.7$\times 10^{38}$ \\
P104 &  704 & PL     & 2.61 &        & 1.6$\times 10^{38}$ \\
P005 &  679 & diskBB & 1.64 & 12.9   & 2.2$\times 10^{38}$ \\
\enddata
\end{deluxetable}

\begin{deluxetable}{lcccc}
\tablecaption{Spectral Change in P098}
\tablehead{ \colhead{Interval} & \colhead{kT} & \colhead{R (km)}
	& \colhead{L$_{\rm (0.2-1.5 keV)}$ (10$^{39}$ ergs\,s$^{-1}$)}
	& \colhead{L$_{\rm bol}$ (10$^{39}$ ergs\,s$^{-1}$)} }
\startdata
a & 0.173$\pm$0.004 & 5200$\pm$100   & 2.04 & 3.2$\pm$0.2 \\
c & 0.123$\pm$0.006 & 9600$\pm$800   & 1.23 & 2.7$\pm$0.5 \\
b & 0.090$\pm$0.007 & 19600$\pm$2000 & 0.78 & 3.3$\pm$0.9 \\
\enddata
\end{deluxetable}

\end{document}